\documentclass[12pt,preprint]{aastex}

\shorttitle{HCN in Comets LINEAR and NEAT}
\shortauthors{Friedel et al.}
\begin{document}

\title{BIMA ARRAY DETECTIONS OF HCN IN COMETS LINEAR (C/2002 T7) AND NEAT (C/2001 Q4)}

\author{D. N. Friedel\altaffilmark{1}, Anthony J. Remijan\altaffilmark{2,3}, L. E. Snyder\altaffilmark{1}, 
M. F. A'Hearn\altaffilmark{4}, Geoffrey A. Blake\altaffilmark{5},
Imke de Pater\altaffilmark{6}, H. R. Dickel\altaffilmark{1}, J. R. Forster\altaffilmark{6}, 
M. R. Hogerheijde\altaffilmark{7}, C. Kraybill\altaffilmark{6}, L. W. Looney\altaffilmark{1}, Patrick Palmer\altaffilmark{8}, \& 
M. C. H. Wright\altaffilmark{6}}

\altaffiltext{1}{Department of Astronomy, 1002 W. Green St., University of
Illinois, Urbana IL 61801\\
email: friedel@astro.uiuc.edu, snyder@astro.uiuc.edu, lwl@astro.uiuc.edu, lanie@astro.uiuc.edu}
\altaffiltext{2}{NASA Goddard Space Flight Center, Computational and Information Sciences and Technologies Office, Code 606
, Greenbelt, MD 20771\\
email: aremijan@pop900.gsfc.nasa.gov}
\altaffiltext{3}{National Research Council Resident Research Associate}
\altaffiltext{4}{Department of Astronomy, University of Maryland, College
Park MD 20742-2421\\
email: ma@astro.umd.edu}
\altaffiltext{5}{Division of Geological and Planetary Sciences; Division of Chemistry and Chemical Engineering 
California Institute of Technology 150-21, Pasadena, CA 91125\\
email: gab@gps.caltech.edu}
\altaffiltext{6}{Department of Astronomy, University of California, Berkeley, CA 94720\\
email: imke@floris.berkeley.edu, rforster@astro.berkeley.edu, ckraybill@astro.berkeley.edu,\\
wright@astro.berkeley.edu}
\altaffiltext{7}{Leiden Observatory, PO Box 9513, 2300 RA, Leiden, The Netherlands\\
email: michiel@strw.leidenuniv.nl}
\altaffiltext{8}{Department of Astronomy and Astrophysics, University of Chicago, Chicago, IL 60637\\
email: ppalmer@oskar.uchicago.edu}

\begin{abstract}
We present interferometric detections of HCN in comets
LINEAR (C/2002 T7) and NEAT (C/2001 Q4) with the Berkeley-Illinois-Maryland
Association (BIMA) Array. With a $25{\farcs}4\times20{\farcs}3$
synthesized beam around Comet LINEAR and using a variable temperature and outflow velocity (VTOV) model, 
we found an HCN column density of $<N_T>=6.4\pm2.1\times10^{12}$~cm$^{-2}$, and a production rate of Q(HCN)=$6.5\pm2.2\times10^{26}$~s$^{-1}$, giving
a production rate ratio of HCN relative to H$_2$O of $\sim3.3\pm1.1\times10^{-3}$ and relative to CN of $\sim4.6\pm1.5$. With 
a $21{\farcs}3\times17{\farcs}5$ synthesized beam around Comet NEAT and using a VTOV model, 
we found an HCN column density
of $<N_T>=8.5\pm4.5\times10^{11}$~cm$^{-2}$, and
a production rate of Q(HCN)=$8.9\pm4.7\times10^{25}$~s$^{-1}$, giving
a production rate ratio of HCN relative to H$_2$O of $\sim7.4\pm3.9\times10^{-4}$ and relative to CN of $\sim0.3\pm0.2$.
For both comets, the production rates relative to H$_2$O are similar to those found in previous comet observations. For Comet LINEAR 
the production rate relative to CN is consistent with HCN being the primary parent species of CN, while
for Comet NEAT it is too low for this to be the case.
\end{abstract}


\section{INTRODUCTION}

HCN has been extensively studied around several comets including Comet P/Halley \citep{desp86}, Comet Hyakutake
(C/1996 B2) \citep[e.g.,][]{mumma96,lis97,woma97,biver99b}, 
Comet Hale-Bopp (C/1995 O1) \citep[e.g.,][]{fitz95,wright98,hir99,irv99,magee99,ziu99,veal00,snyder01,wood02}
and Comet LINEAR (C/1999 S4) \citep{b-m01,hoger04}.
\citet{veal00} made some of the first interferometric observations of HCN in 
Comet Hale-Bopp. From the distribution and temporal behavior of the HCN emission,
they were able to make very accurate calculations of the HCN production rate as
well as modeling its distribution in the coma using a spherical Haser model \citep{haser}.
The deviations from the Haser model were explained by the existence of jets releasing
HCN gas, a conclusion directly supported by high spatial resolution interferometric imaging of Comet Hale-Bopp by \citet{blake99}.
\citet{snyder01} followed up the observations of \citet{veal00}
and found the measured HCN scale length to be very similar to the theoretical 
predictions of \citet*{heub92a} and \citet{crov94}.

In the Spring of 2004, we were awarded a rare opportunity to observe two comets
passing into the inner solar system and within $\sim$0.3 AU of the Earth. Comet
LINEAR (C/2002 T7) reached perigee on 2004 May 19, coming within 
0.27 AU, while Comet NEAT (C/2001 Q4) reached perigee on 2004 May 7, 
coming within 0.32 AU. Both comets were 
identified as being dynamically new; such comets reach a peak of strong activity as 
they approach perihelion. It was extremely important to observe both comets 
when they were close to both perihelion and perigee in order to observe them near their peak emission
and their maximum signal due to their proximity.

This paper describes the results of an effort to observe HCN in both Comets
LINEAR and NEAT, with the Berkeley-Illinois-Maryland Association (BIMA) Array\footnote{Operated by the
University of California, Berkeley, the University of Illinois, and the
University of Maryland with support from the National Science Foundation.}
near Hat Creek, California ($\S$2). We present interferometric
detections of HCN in both comets and calculate the total beam averaged
HCN column densities and production rates by two methods. The first assumes the realistic scenario of a temperature and outflow velocity 
that vary with cometocentric distance and the second model assumes a constant temperature and outflow velocity throughout the coma.

\section{OBSERVATIONS}

We used the BIMA Array
at Hat Creek, California (122$^o$28$'$8$''$.4 West, 40$^o$49$'$2$''$.28 North; altitude
1333 m) in D-configuration (baselines from $\sim$6m to $\sim$35m) cross-correlation mode to observe
HCN in Comets LINEAR (C/2002 T7) and NEAT (C/2001 Q4) during their 2004 apparitions. 

The HCN observations of Comet LINEAR, using JPL reference orbit 69, 
were taken 2004 May 11 toward the topocentric coordinates $\alpha$=01$^h$15$^m$59.$^s$06; 
$\delta$=$-$07$^o$39$'$16.5$''$ [J2000.0]\footnote{We believe the positional accuracy to 
be good to a few arcseconds for our observations. This error is insignificant relative to the size of the synthesized beams (see $\S$\ref{sec:res}).} at the beginning of our observations (16:54 UT), and moved to $\alpha$=01$^h$18$^m$20.$^s$56; 
$\delta$=$-$07$^o$52$'$33.9$''$ [J2000.0] by the end of our observations (20:58 UT).
The comet was at a heliocentric distance of 0.73 AU and a geocentric distance
of 0.44 AU (1\arcsec=319 km at 0.44 AU). The spectral window containing the HCN hyperfine components had a bandwidth of 50 MHz
and was divided into 128 channels giving a channel spacing of 0.39 MHz (1.321 km s$^{-1}$).
W3(OH) was used as the flux density calibrator for this observation. The quasar
0108+015 was used to calibrate the antenna based gains.

The HCN observations of Comet NEAT, using JPL reference orbit 123, 
were taken 2004 May 23 toward the topocentric coordinates $\alpha$=09$^h$13$^m$54.$^s$67; 
$\delta$=36$^o$46$'$30.7$''$ [J2000.0] at the beginning of our observations (20:58 UT), and moved to 
$\alpha$=09$^h$15$^m$23.$^s$16; $\delta$=37$^o$18$'$34.9$''$ [J2000.0] by the end of our observations (06:51 May 24 UT).
The comet was at a heliocentric distance of 0.97 AU and a geocentric distance
of 0.61 AU (1\arcsec=442 km at 0.61 AU). The spectral resolution was the same as for Comet LINEAR.
Mars was used as the flux density calibrator and 0927+390 was used to calibrate the 
antenna based gains.

 The absolute amplitude
calibration is accurate to within $\sim$20\%. We note that our observations may cover a significant amount 
of the rotation period of Comets LINEAR and NEAT. However, to date no rotation period has been given 
for either comet, thus we are unable to comment on any effects this may have on our results. The data 
were combined and imaged using the MIRIAD software package \citep{sault95}.
Table~\ref{tab:hcn} lists the HCN molecular line parameters. The HCN spectroscopic constants were taken from \citet{maki74}.
In the following, our analysis is for the strongest component, the $J=1-0,$ $F=2-1$ transition.

\section{RESULTS}\label{sec:res}

Figure~\ref{fig:spec}(a-c) shows our map and spectra of HCN around Comet LINEAR
(C/2002 T7). Figure~\ref{fig:spec}a shows the map of the $J=1-0, F=2-1$ transition of HCN in 1 $\sigma$ contours, starting at 2 $\sigma$.
The synthesized beam of $25\farcs 4\times20\farcs 3$ is shown at the bottom left of
the map. The line segment in the image shows the projection of the direction toward the sun.
The origin of the line segment is anchored at the predicted position of the comet nucleus. 
The coordinates are given in offset arcseconds centered on the comet nucleus.
We note that the peak intensity of the HCN emission is offset from the predicted position of the nucleus by $\sim$5\arcsec\ along the semi-major 
axis of the synthesized beam. In general, the positional uncertainty roughly scales as the beamsize/SNR. For Comet LINEAR
this uncertainty is 4-5\arcsec, so the offset is not significant\footnote{This does not include any uncertainties in the ephemeris.}.
Figure~\ref{fig:spec}b shows the cross-correlation spectrum 
of this transition. The dashed line corresponds to the rest frequency of the $J=1-0, F=2-1$ line for a 
cometocentric rest velocity of 0 km s$^{-1}$. The 1 $\sigma$ rms noise level is 
seen at the left of the panel. The $F=2-1$ line for Comet LINEAR was fit with a Gaussian by a least-squares method which gives a peak intensity
of 0.59$\pm$0.14 Jy beam$^{-1}$, a FWHM of 1.38$\pm$0.34 km s$^{-1}$, and a cometocentric velocity of -0.29$\pm$0.28 km s$^{-1}$.
Figure~\ref{fig:spec}c shows the cross-correlation spectrum 
(Hanning smoothed over three channels) of this transition.

Figure~\ref{fig:spec}(d-f) shows our map and spectra of HCN around Comet NEAT
(C/2001 Q4). Figure~\ref{fig:spec}d shows the map of the $J=1-0, F=2-1$ transition of HCN in 1 $\sigma$ contours, starting at 2 $\sigma$.
The synthesized beam of $21\farcs 3\times17\farcs 5$ is shown at the bottom left of
the map. We note, as with Comet LINEAR, that the HCN peak intensity is offset by $\sim$5\arcsec\ along the
semi-major axis of our synthesized beam and that there is also some elongation of the source along this axis.
We consider neither the offset nor the elongation to be significant because these are most likely due to the beam 
size and the low SNR.
 Figure~\ref{fig:spec}e shows the cross-correlation spectrum 
of this transition. The dashed line is similar to Figure~\ref{fig:spec}b for a 
cometocentric rest velocity of 0 km s$^{-1}$. The $F=2-1$ line appears in only a single channel so
we will assume a FWHM of $\sim$1.3 km s$^{-1}$ for HCN.
Figure~\ref{fig:spec}f shows the cross-correlation spectrum 
(Hanning smoothed over three channels) of this transition.

\section{DISCUSSION}
\subsection{Column Densities and Production Rates}\label{sec:NT}
In the innermost coma, collisions between cold cometary species dominate 
the rotational population of the gas, but in the less dense outer coma 
the collision time becomes longer than the time for absorption of solar photons and the subsequent cascade of emission
In that case, the population can also be 
determined by solar radiation driven fluorescence. Fluorescence effects 
could allow LTE to be maintained at a lower density than required in 
optically thin cases \citep[e.g.,][]{crov84,b-m87}. \citet{lovell04}
determined that for the $J=1-0$ transition of HCN for Comet Hyakutake (C/1996 B2) the transition between collisionally dominated and fluorescence dominated emission
is between 26,000 and 38,000 km from the nucleus. Since both Comet LINEAR and Comet NEAT have similar water production
rates relative to Comet Hyakutake (D.~G.~Schleicher, 2004, private communication)(see \S\ref{sec:h2o}) at comparable heliocentric distances, we assume their transition
radii are also comparable. On the sky this translates to 82-119\arcsec\ for Comet LINEAR and 59-86\arcsec\ 
for Comet NEAT.

The synthesized beams sample a cylinder of gas through each comet\footnote{In a small cylinder passing through the cometary gas, only a small fraction of the HCN will be in the less dense fluorescent zone. Hence, unlike the case for some optical molecular observations, here we assume that the HCN in the fluorescence zone can be ignored relative to that in the denser collisional zone.}, and because the HCN volume density falls off faster than $r^{-2}$ due to photodissociation,
most of the contribution to the integrated line intensity will come from a sample on the order of the size of the projected synthesized beam. Since the synthesized
beam is much smaller than the outer radii of the collisionally dominated regions, conditions significantly outside of the beam are not likely to affect our observations. 

The contribution from different cometocentric radii (i.e. from different source sizes) is determined by two factors. The first is that sources of differing size but 
equal total flux do not have equal peak fluxes and thus, when summed together, do not contribute equally to the total peak flux of a map. The second is that the array 
starts to resolve out flux from sources that are much larger than the synthesized beam. The first effect is dominant for source sizes that are on the order of the 
synthesized beam while the second becomes more dominant for larger source sizes. These two effects are illustrated in Figure~\ref{fig:resp}. The top panel is from 
our Comet LINEAR observations and the lower panel is from our Comet NEAT observations. The abscissa is cometocentric radius in arcseconds of a source on the sky 
and the ordinate is $r_i$, the array recovery factor (see Appendix~\ref{adx:model} for a detailed explanation). The collisionally dominated region of the coma is 
labeled ``Collisional Zone'', the transition between the collisionally dominated and fluorescence dominated regions is denoted by the hatching, and the fluorescence 
dominated region is labeled ``Fluorescence Zone''. The solid line shows the peak flux detected by the array from each source size as a percentage of the peak flux 
of a point source of equal total flux. Thus, both from this figure and the above argument, it is clear that we are sampling the collisionally dominated region and 
are insensitive to fluorescence effects.

Our calculations are made assuming optically thin emission, since the limits on the satellite hyperfine components 
show that they are significantly weaker than the main component. For cross-correlation observations,
$<N_T>$, the beam-averaged molecular column density, is 
\begin{equation}
<N_T> = {2.04~W_{\rm I}~Z~e^{E_{\rm u}/T_{\rm rot}}\over \theta_{\rm a}\theta_{\rm b} S \mu^2 \nu^3} \times~10^{20}~{\rm cm}^{-2}.\label{eq:NTOT}
\end{equation}
In equation~(\ref{eq:NTOT}), $W_{\rm I} = \int I_\nu dv$ in Jy beam$^{-1}$ km s$^{-1}$, 
and $I_\nu$ is the flux density per beam. $Z$ is the rotational partition 
function ($Z=0.47T_{\rm rot}$), $E_u$ is the upper state energy of the transition, $T_{\rm rot}$ is the rotation temperature,
 $\theta_{\rm a}$ and $\theta_{\rm b}$ are the FWHM synthesized 
beam dimensions in arcsec, $S$ is the line strength, $\mu$ the
dipole moment in Debye, and $\nu$ is the frequency in GHz.

 From the least squares fit to the Comet LINEAR spectrum, we find an integrated line 
intensity of 1.46$\pm$0.49 Jy beam$^{-1}$ km s$^{-1}$ for the $J=1-0, F=2-1$ HCN line around Comet
LINEAR\footnote{To convert this to integrated main beam brightness temperature multiply by 0.30 K/(Jy beam$^{-1}$).}.
From Figure~\ref{fig:spec}e, we find an integrated line intensity of 0.22$\pm$0.11
Jy beam$^{-1}$ km s$^{-1}$ for the $J=1-0, F=2-1$ HCN line around Comet
NEAT\footnote{To convert this to integrated main beam brightness temperature multiply by 0.42 K/(Jy beam$^{-1}$).}.

\subsubsection{Model 1: Assuming Variable Temperatures and Outflow Velocities}\label{sec:non}
We consider the case where the temperature and outflow velocity 
 vary with cometocentric distance, as has been observed and modeled in previous comets (e.g. Comet Hyakutake and Comet Hale-Bopp \citep{combi99,combi99b}). 
We label this as the variable temperature and outflow velocity (VTOV) model. In order to do this, we modeled
the HCN emission as concentric shells (of thickness 100 km out to a radius of 10,000 km and a thickness of 2000 km for 
radii of 10,000-30,000 km)
around the nucleus and calculated the total column density and production rate.
Since we consider both comets to be similar to Comet Hyakutake, we used the
velocity profiles from Figure 3 of \citet{combi99} or the fit to that data from \citet{lovell04} to give velocities to each shell. The
April 9 profile was used for Comet LINEAR and the March 30 profile and fit was used for Comet NEAT, since on these dates Comet Hyakutake was at a similar
heliocentric distance. These velocities were scaled to match the
observed velocity profile of the spectral line, taking into account thermal broadening. Similarly, the temperature profiles
for each comet were also taken from \citet{combi99}. Our VTOV model is explained in detail in Appendix~\ref{adx:model}, and gives total beam averaged column densities of
$<N_T>$ = 6.4$\pm$2.1$\times~10^{12}$ cm$^{-2}$ for Comet LINEAR
and $<N_T>$ = 8.5$\pm$4.5$\times~10^{11}$ cm$^{-2}$ for Comet NEAT and production rates of
Q(HCN)=6.5$\pm2.2\times10^{26}$ s$^{-1}$ for Comet LINEAR and
Q(HCN)=8.9$\pm4.7\times10^{25}$ s$^{-1}$ for Comet NEAT.

\subsubsection{Model 2: Assuming Constant Temperatures and Outflow Velocities}\label{sec:haser}
In the case of assuming a constant $T_{rot}$ the calculation of total beam averaged column density is straightforward, using equation~(\ref{eq:NTOT}). 
For Comet LINEAR we assume a rotation temperature of 115 K \citep{disanti04,magee04,kuppers04} and for Comet NEAT we 
assume a rotation temperature of 52 K. The temperature for Comet NEAT was derived from Figure 3 of \citet{combi99} by averaging across the section of their March 30 curve 
we are sensitive to, when Comet Hyakutake was at a similar heliocentric distance. 
Using the above temperature for Comet LINEAR we find a total beam averaged HCN column density (corrected for the other hyperfine components) 
of $<N_T>$ = 9.5$\pm$3.2$\times~10^{12}$ cm$^{-2}$. Using the above temperature for Comet NEAT we find a total beam averaged HCN column density 
(corrected for the other hyperfine components) of $<N_T>$ = 9.6$\pm$5.1$\times~10^{11}$ cm$^{-2}$. 

To calculate $Q_p$, the parent molecule production rate, we use the total beam averaged column densities with the Haser model \citep{haser}.
Since the Haser model assumes a constant outflow velocity, it is better suited for the fluorescence region rather than the inner, collisionally dominated, coma.
However, we will compare the Haser model results with our model since the Haser model has been commonly used by previous cometary observations \citep{hoger04,snyder01,veal00,wright98}.

For a photodissociation scale length $\lambda _p$, nuclear radius $r_n$, and 
constant radial outflow velocity $v_0$, the density as a function of 
$r$, $n_p(r)$, is given by \citet{snyder01}:
\begin{equation}
n_p(r) = \frac{Q_p}{4 \pi r^2 v_0} e^{- \frac{(r-r_n)}{\lambda _p}}.\label{eq:np}
\end{equation}

For Comets LINEAR and NEAT, we assume a nuclear radius
of $r_n$ = 5 km. We use a radial outflow velocity of $v_0$ = 
0.62 km s$^{-1}$ for Comet LINEAR, based on the HWHM of the Gaussian fit and taking into account the 10\% 
increase in line width due to thermal broadening \citep{biver99a} and $v_0$ = 0.59 km s$^{-1}$ for Comet NEAT, based on
the single channel half width of the line and taking into account thermal broadening.

Consider a point on the line of sight that passes a projected distance, $p$,
in the plane of the sky from the nucleus. For a given value of $z$, measured 
along the line of sight, this will correspond to a radial distance 
$r=\surd (p^2 + z^2)$ from the nucleus. The column density, $N_p$, is given by 
\citet{snyder01}:
\begin{equation}
N_p(p) = \frac{Q_p}{4 \pi v_0} e^{\frac{r_n}{\lambda _p}}\int_{-\infty}^{\infty}\frac{{\rm exp}\{-[(p^2+z^2)^{1/2}/\lambda_p]\}}{(p^2+z^2)} dz.\label{eq:Npp}
\end{equation}
Finally, $N_p$ from equation~(\ref{eq:Npp}) may be averaged over the synthesized
beam and equated to $<N_T>$, the beam-averaged molecular column density,
in order to obtain $Q_p$, the parent molecule production rate from the Haser 
model.

The Quiet Sun HCN photodissociation rate in the Solar UV field at $r_{hel}$ 
= 1 AU is given by \citet{heub92a} as $\alpha$(1 AU) = 
1.3 $\times ~10^{-5}$ s$^{-1}$ and by \citet{crov94} as 
1.5 $\times ~10^{-5}$ s$^{-1}$. The expected accuracy is 10-20\% \citep{crov94}, 
so this is reasonable agreement. Thus we will assume $\alpha$(1 AU)=1.3 $\times ~10^{-5}$ s$^{-1}$
(equivalent to ${\sim}3.7{\times}10^4$ km for Comet NEAT 
and ${\sim}2.2{\times}10^4$ km for Comet LINEAR).
We find that the May 11 data around Comet LINEAR
give a production rate, Q(HCN), of 5.1$\pm$1.7$\times~10^{26}$ s$^{-1}$, and 
the May 23 data around Comet NEAT give a production rate of 5.7$\pm$3.0$\times~10^{25}$ s$^{-1}$.

\subsubsection{Comparison of Models}\label{sec:comp}
The column densities calculated with the VTOV model (model 1) are 11-33\% lower than those calculated with model 2. However, the production rates calculated
with the VTOV model are 22-36\% higher than those calculated with the Haser model (see model 2 in $\S$\ref{sec:haser}).
These differences are not large. However, the VTOV model provides a better approximation than the Haser model for the physical conditions in the cometary collisional regions that the BIMA Array sampled. Consequently, for the interferometric observations discussed in this paper, the VTOV model should produce more realistic results than the Haser model.
Table~\ref{tab:comp} gives a comparison of the two methods. Column 1 gives the comet name, column 2 gives the total column density,
column 3 gives the production rate, column 4 gives the HCN/H$_2$O ratio (see \S\ref{sec:h2o}) and column 5 gives the HCN/CN ratio (see \S\ref{sec:cn}).

\subsection{Relative Production Rates of HCN to H$_2$O}\label{sec:h2o}
\subsubsection{Comet NEAT}
D.~G.~Schleicher (2005, private communication) measured H$_2$O production rates from Comet NEAT and derived a scaling law of $Q(H_2O)\sim r_H^{-4.3}$, where $r_H$ is the 
heliocentric distance of the comet. From his observations in May and June of 2004 and the scaling law, we scaled the H$_2$O production rates from the heliocentric distances when they were measured to 0.97 
AU for Comet NEAT to match the $r_H$ of our observations. Table~\ref{tab:h2o} gives the measurements and their scaled values.
Column 1 gives the date of observation, column 2 gives the heliocentric distance of the comet, columns 3 and 4 give the measured 
production rates of H$_2$O and CN (see \S\ref{sec:cn}), and columns 5 and 6 give the scaled production rates for H$_2$O and CN, 
respectively. The average value for Comet NEAT is Q(H$_2$O)$\sim1.2\times10^{29}$~s$^{-1}$, which is very similar to water 
production rates measured from Comet Hyakutake at similar heliocentric distances to our observations ($\sim1\times10^{29}$ s$^{-1}$ \citep{lis97}).
If we assume a constant temperature and outflow velocity we find the production rate of HCN relative to H$_2$O for Comet NEAT is $\sim4.7\pm2.5\times10^{-4}$. On the other hand, if we assume the more realistic VTOV model, we
find the production rate of HCN relative to H$_2$O to be $\sim7.4\pm3.9\times10^{-4}$ for Comet NEAT.
This is similar to the ratios observed around previous comets such as Comet Hyakutake ($1\times10^{-3}$) \citep{lis97}. 

\subsubsection{Comet LINEAR}
H$_2$O production rates were also measured for Comet LINEAR by D.~G.~Schleicher (2005, private communication). From his data, he estimated
the H$_2$O production rate to be $\sim2\times10^{29}$ s$^{-1}$ during our May 11 observations. If we assume a constant temperature and outflow 
velocity, we find the production rate of HCN relative to H$_2$O is $\sim2.6\pm0.9\times10^{-3}$ for Comet LINEAR. On the other hand, if we assume the more realistic VTOV model, we find the production rate of HCN relative to H$_2$O to be $\sim3.3\pm1.1\times10^{-3}$.
This is similar to the ratio observed around Comet Hale-Bopp ($2.1-2.6\times10^{-3}$) \citep{snyder01}.

\subsection{Relative Production Rates of HCN to CN}\label{sec:cn}
\subsubsection{Comet NEAT}
D.~G.~Schleicher (2005, private communication) measured CN production rates from Comet NEAT and derived a scaling law of
$Q(CN)\sim r_H^{-2.1}$. From his observations in May and June 2004 and the scaling law, we scaled 
these values from the heliocentric distances when they were measured to the
heliocentric distance of Comet NEAT during our observations (see Table~\ref{tab:h2o}). Averaging these values 
gives Q(CN) $\sim2.6\times10^{26}$ s$^{-1}$ for Comet NEAT.
If we assume a constant temperature and outflow velocity we find the production rate of HCN relative to CN is $\sim0.2\pm0.1$ for Comet NEAT. 
On the other hand, if we assume the more realistic VTOV model, we
find the production rate of HCN relative to CN of $\sim0.3\pm0.2$ for Comet NEAT.
The photodissociative branching ratio of HCN implies that $\sim$97\% of HCN will be photodissociated
into H and CN \citep*{heub92b}. \citet{wood02} concluded that HCN is most likely the primary parent species of CN.
Thus, one would expect the HCN to CN ratio to be near 1.
However, the ratio for Comet NEAT is a factor of $\sim$3 smaller than the expected ratio.

\subsubsection{Comet LINEAR}
CN production rates were measured for Comet LINEAR by D.~G.~Schleicher (2005, private communication). From his data, he estimated
the CN production rate to be $\sim1.4\times10^{26}$ s$^{-1}$ during our May 11 observations. If we assume a constant temperature and outflow 
velocity we find the production rate of HCN relative to CN is $\sim3.6\pm1.2$ for Comet LINEAR. On the other hand, if we assume the more realistic VTOV model, we find the production rate of HCN relative to CN to be $\sim4.6\pm1.5$.

\section{SUMMARY}
We have detected the $J=1-0$ transition of HCN in Comets NEAT (C/2001 Q4) and LINEAR (C/2002 T7).
We have calculated the total column density and production rates by two different models. The first (VTOV) model assumes 
the realistic scenario of temperature and outflow velocity that vary with cometocentric distance. We compare this model to
one where the temperature and outflow velocity are constant throughout the coma (Haser model). The differences between
the outcomes of the models are $\sim$11-33\% for $N_T$ and $\sim$22-36\% for $Q$. 
However, for the interferometric observations described in this paper,
more realistic results will be obtained by using the VTOV model outlined in Sec.~\ref{sec:non} and Appendix~\ref{adx:model}.
This model gives production rates of 
Q(HCN)=$6.5\pm2.2\times~10^{26}$~s$^{-1}$ and ratios of HCN/H$_2$O and HCN/CN of $\sim3.3\pm1.1\times10^{-3}$ and $\sim4.6\pm1.5$, respectively, for Comet LINEAR and 
Q(HCN)=$8.9\pm4.7\times~10^{25}$~s$^{-1}$ and ratios of HCN/H$_2$O and HCN/CN of $\sim7.4\pm3.9\times10^{-4}$ and $\sim0.3\pm0.2$, respectively, for Comet NEAT.

The HCN production rate relative to H$_2$O for Comet NEAT is similar to that found previously for Comet Hyakutake \citep{lis97}, 
while the ratio for Comet LINEAR is similar to that observed from the highly productive Comet Hale-Bopp
\citep{snyder01}. The HCN production rate relative to CN for Comet LINEAR is consistent with HCN being the
primary parent of CN, as was suggested in \citet{wood02}. However, for Comet NEAT the value is too low.

\acknowledgements
We thank J.~R. Dickel for assisting with the observations, an anonymous referee for many helpful
comments, and D.~G. Schleicher for providing H$_2$O and CN production rates. 
This work was partially funded by: NSF AST02-28953, AST02-28963, AST02-28974 and AST02-28955; and the Universities of Illinois, Maryland, and California, Berkeley.

\bigskip
\appendix
\centerline{APPENDIX}
\section{DESCRIPTION OF THE VTOV MODEL}\label{adx:model}
The variable temperature and outflow velocity (VTOV) model consists of concentric spherical shells centered on a cometary nucleus of radius 5 km. The shells had a thickness
of 100 km (with the exception of the 1st shell (95 km)) out to a cometocentric radius of 10,000 km and a thickness of 2,000 km for radii from 10,000 to 30,000 km. 30,000 
km is equivalent to a cometocentric radius of 94\arcsec on the sky for Comet LINEAR and 67.9\arcsec for Comet NEAT. This distance was chosen as the outer 
limit of our model because the structures of this size contribute less than 0.1\% to the peak flux and the population of HCN
at this radius is less than 40\% that of the central shell due to photodissociation. Thus, by the combination of these factors
the contribution from any shells beyond 30,000 km would be negligible.

While quantities such as the outflow velocity and temperature vary in a very non-linear way across the coma of each comet, the
changes across any given shell are small enough to be considered linear. Thus, the quantities calculated for each shell, unless
otherwise specified, are calculated at the average radius of each shell, since they will be very close to the average value
for the entire shell.

For the model we calculate the following quantities for each shell $i$ in our model :
\begin{enumerate}
\item The time, $t_i$, each molecule spends in the shell, is given by\label{it:ti}
\begin{equation}
t_i=\frac{(R_i - R_{i-1})}{V_i},\label{eqn:ti}
\end{equation}
where $R_i$ and $R_{i-1}$ are the outer and inner radii of shell $i$, respectively, and $V_i$ is the average velocity of the 
molecule across the shell (from \citet{combi99}).
\item The total time, $t_{i,\rm tot}$, it takes a molecule to travel from the nucleus to the center of the shell, is given by\label{it:ttot}
\begin{equation}
t_{i,\rm tot}=\left( \sum_{j=1}^{i-1}t_j\right) + \frac{t_i}{2}.\label{eqn:ttot}
\end{equation}
\item The number of molecules, $n_i$, in each shell, is given by\label{it:num}
\begin{equation}
n_i=Qt_ie^{-\alpha t_{i,\rm tot}/r_H^2},\label{eqn:num}
\end{equation}
where $Q$ is the production rate, $\alpha$ is the photodissociation rate (assumed to be $\alpha$(1 AU)=1.3 $\times ~10^{-5}$ 
s$^{-1}$), and $r_h$ is the heliocentric radius of the comet. {\it Note}: At this point it does not matter what value we give $Q$ since it does not vary from shell to 
shell and thus is a constant for this part of the model.
\item The BIMA Array's flux recovery factor for the projected size of the shell on the sky is defined as 
the peak flux of a shell compared to that of a point source. This is a combination of a geometric effect (i.e. the flux is spread out over larger 
areas for larger shells), which is dominant for source sizes comparable to our synthesized beam size, and the fact that the array 
resolves out flux from sources that are much larger than the synthesized beam (array filtering effect). 
The recovery factor is found in the following way:
\begin{enumerate}
\item Each source size (or shell) is modeled by first creating a Gaussian\footnote{In this step we approximate the 
shells as Gaussian sources rather than as actual 
shells because any structure on the sky that has sharp edges, such as a shell, will produce ringing in the $u-v$ plane and 
artifacts in the final map. Since the actual observations of the comets observed continuous distributions on the sky with no 
sharp edges, approximating the shell by a Gaussian is reasonable.} source image for each concentric shell,
 of the appropriate FWHM and at the position on the sky of the comet, with the MIRIAD task IMGEN\footnote{Detailed
descriptions of all MIRIAD tasks can be found at: http://bima.astro.umd.edu/miriad/ref.html}. 
Every Gaussian shell is assumed to contribute equally to the flux recovery for a uniformly extended source.
\item In order to remove geometric and filtering effects each image is run through UVMODEL to produce a $u-v$ data set that is the array's response to the input source based on the $u-v$ 
tracks of our actual observations. 
\item Each data set is then INVERTed, CLEANed, and RESTORed as normal to produce the final output map. 
\item The peak intensity ($p_i$) of each map is compared to that of a point source ($p_{\rm ps}$) model of equal integrated intensity to give the 
array's recovery factor, $r_i$, where,
\begin{equation}
r_i=\frac{p_i}{p_{\rm ps}}.
\end{equation}
Note that $r_i{\rightarrow}0$ for an infinitely extended source, $r_i{\rightarrow}1$ for a point source, and $r_i=0.5$ if the source just fills the beam.
\end{enumerate}The array's recovery factor is 
illustrated in Figure~\ref{fig:resp}. The top panel is for Comet LINEAR and the bottom panel is for Comet NEAT. The abscissa 
is cometocentric radius and the ordinate is $r_i$. The solid line shows $r_i$ for all shells.\label{it:resp}
\item The production rate $Q$ of the comet is found by modifying equation~(\ref{eq:NTOT}),
\begin{equation}
\frac{4ln(2)n_ir_i}{\pi ab}=<N_i>=\frac{2.04 W_i Z_i e^{E_u/T_{i}}}{\theta_{\rm a}\theta_{\rm b}S\mu^2\nu^3}\times10^{20} {\rm cm}^{-2}.\label{eqn:Ni}
\end{equation}
We note that $4ln(2)/\pi ab$ is the projected area on the sky of our Gaussian synthesized beam (in cm$^2$), $N_i$ is the beam averaged column density of the shell, $W_i$ is the
integrated line intensity of the shell, and $T_i$ 
is the rotation temperature for the shell (from \citet{combi99}). Rearranging equation~(\ref{eqn:Ni}) and incorporating the fact that
\begin{equation}
W=\sum_{i}W_i
\end{equation}
gives
\begin{equation}
W=\frac{1.36\theta_a\theta_b S\mu^2\nu^3}{\pi ab}\sum_{i}\frac{n_ir_i}{Z_ie^{E_u/T_{i}}}\times10^{-20}\label{eqn:sum1}.
\end{equation}
Incorporating equation~(\ref{eqn:num}) and rearranging equation~(\ref{eqn:sum1}) gives
\begin{equation}
Q=\frac{0.736W\pi ab}{\theta_a\theta_b S\mu^2\nu^3}\left(\sum_{i}\frac{r_it_ie^{-\alpha t_{i,\rm tot}/r_H^2}}{Z_ie^{E_u/T_{i}}}\right)^{-1}\times10^{20}.\label{eqn:sum2}
\end{equation}
Since all factors on the right hand side of equation~(\ref{eqn:sum2}) are known, $Q$ can be found.
{\it Note}: In equation~(\ref{eqn:Ni}) we divided $n_i$ by the projected beam area rather than the projected area of the shell, even for those shells with a projected area larger than our beam. This is done because, even though there are some molecules outside of the synthesized beam, they are contributing flux to the area inside the synthesized beam.
\item Now that $Q$ has been found, use it in equation~(\ref{eqn:num}) to find $n_i$. Then use each $n_i$ in equation~(\ref{eqn:Ni})
to find $N_i$ for each shell. Then the total beam averaged column density, $<N_T>$, is given by the sum over all shells,
\begin{equation}
<N_T>=\sum_{i}<N_i>.
\end{equation}
\end{enumerate}

\clearpage
\begin{deluxetable}{cccc}
\tabletypesize{\scriptsize}
\tablewidth{20pc}
\tablecolumns{4}
\tablecaption{HCN Molecular Line Parameters}
\tablehead{
\colhead{Quantum} & \colhead{Frequency} & \colhead{E$_{u}$} & \colhead{$<S_{i,j}\mu^{2}>$}\\
\colhead{Numbers} & \colhead{(MHz)} & \colhead{(K)} & \colhead{(Debye$^{2}$)}
}
\startdata
$J=1-0, F=1-1$ & 88,630.4157(10) & 4.3 & 3.0\\
$J=1-0, F=2-1$ & 88,631.8473(10) & 4.3 & 4.9\\
$J=1-0, F=0-1$ & 88,633.9360(10) & 4.3 & 1.0\\
\enddata
\label{tab:hcn}
\end{deluxetable}

\begin{deluxetable}{cccccc}
\tabletypesize{\scriptsize}
\tablewidth{0pt}
\tablecolumns{6}
\tablecaption{Comparison of Constant Temperature and Velocity Results with Variable Temperature and Outflow Velocity (VTOV) Results}
\tablehead{
\colhead{} & \colhead{$<N_T>$} & \colhead{Q(HCN)} & \colhead{Q(HCN)} & \colhead{} & \colhead{Q(HCN)}\\
\cline{4-4} \cline{6-6}
\colhead{Comet} & \colhead{(cm$^{-2}$)} & \colhead{(s$^{-1}$)} & \colhead{Q(H$_2$O)} & \colhead{} & \colhead{Q(CN)}}
\startdata
\cutinhead{Constant Temperature and Outflow Velocity (Haser model)}
LINEAR & $9.5\pm3.2\times10^{12}$ & $5.1\pm1.7\times10^{26}$ & $2.6\pm0.9\times10^{-3}$ & & $3.6\pm1.2$\\
NEAT & $9.6\pm5.1\times10^{11}$ & $5.7\pm3.0\times10^{25}$ & $4.7\pm2.5\times10^{-4}$ & & $0.2\pm0.1$\\
\cutinhead{Variable Temperature and Outflow Velocity (VTOV model)}
LINEAR & $6.4\pm2.1\times10^{12}$ & $6.5\pm2.2\times10^{26}$ & $3.3\pm1.1\times10^{-3}$ & & $4.6\pm1.5$ \\
NEAT & $8.5\pm4.5\times10^{11}$ & $8.9\pm4.7\times10^{25}$ & $7.4\pm3.9\times10^{-4}$ & & $0.3\pm0.2$\\
\enddata
\label{tab:comp}
\end{deluxetable}

\begin{deluxetable}{cccccc}
\tabletypesize{\scriptsize}
\tablewidth{0pt}
\tablecolumns{6}
\tablecaption{Scaling of H$_2$O and CN Production Rates From Comet NEAT\tablenotemark{a}}
\tablehead{
\colhead{} & \colhead{$r_H$} & \colhead{Q(H$_2$O)} & \colhead{Q(CN)} & \colhead{Scaled Q(H$_2$O)\tablenotemark{b}} & \colhead{Scaled Q(CN)\tablenotemark{c}}\\
\colhead{Date} & \colhead{(AU)} & \colhead{(s$^{-1}$)} & \colhead{(s$^{-1}$)} & \colhead{(s$^{-1}$)} & \colhead{(s$^{-1}$)}}
\startdata
2004 May 11 & 0.97 & $1.3\times10^{29}$ & $2.6\times10^{26}$ & $1.3\times10^{29}$ & $2.6\times10^{26}$\\
2004 May 12 & 0.96 & $1.3\times10^{29}$ & $2.5\times10^{26}$ & $1.2\times10^{29}$ & $2.4\times10^{26}$\\
2004 June 09 & 1.05 & $8.9\times10^{28}$ & $2.4\times10^{26}$ & $1.3\times10^{29}$ & $2.8\times10^{26}$\\
2004 June 10 & 1.06 & $7.9\times10^{28}$ & $2.1\times10^{26}$ & $1.2\times10^{29}$ & $2.6\times10^{26}$\\
\enddata
\tablenotetext{a}{Unscaled data from D. G. Schleicher(private communication).}
\tablenotetext{b}{The H$_2$O production rates were scaled by $r_H^{-4.3}$ for each comet (see$\S$\ref{sec:h2o}).}
\tablenotetext{c}{The CN production rates were scaled by $r_H^{-2.1}$ for each comet (see$\S$\ref{sec:cn}).}
\label{tab:h2o}
\end{deluxetable}

\clearpage
\figcaption{Comet LINEAR (C/2002 T7) and NEAT (C/2001 Q4) single field HCN images and spectra. (a) Comet T7 emission contours from the $J=1-0, F=2-1$ transition of HCN at 88.6318 GHz. Contours indicate the HCN emission near its peak centered at a cometocentric velocity of 0 km/s. The contour levels are -0.226, 0.226, 0.339, 0.452, and 0.565 Jy/beam ($1\sigma$ spacing). The peak is 0.63 Jy/beam. Image coordinates are arcseconds offsets relative to the predicted position of the nucleus. The synthesized beam of $25.41{\arcsec}\times20.34\arcsec$ is in the lower left and the line segment shows the solar direction. (b) HCN cross-correlation spectra. Abscissa is radial velocity relative to the comet nucleus. Ordinate is flux density per beam, I$_\nu$, in Jy/beam; $\sigma\sim$ 0.113 Jy/beam (indicated by the vertical bar at the left). The dashed line is centered on the rest frequency (88.6318 GHz).(c) HCN cross-correlation spectra (Hanning smoothed over 3 channels), labels the same as in (b). (d) Comet Q4 emission contours from the $J=1-0, F=2-1$ transition of HCN. Contours indicate the HCN emission near its peak centered at a cometocentric velocity of 0 km/s. The contour levels are -0.06, 0.06, 0.09, 0.12 and 0.15 Jy/beam ($1\sigma$ spacing). The peak is 0.17 Jy/beam. Image coordinates are the same as in (a). The synthesized beam of $21.27{\arcsec}\times17.53\arcsec$ is in the lower left and the line segment shows the solar direction. (e) HCN cross-correlation spectra, abscissa and ordinate are the same as in (b); $\sigma\sim$ 0.03 Jy/beam (indicated by the vertical bar at the left). The dashed line is centered on the rest frequency (88.6318 GHz). (f) HCN cross-correlation spectra (Hanning smoothed over 3 channels), abscissa and ordinate are the same as in (b).\label{fig:spec}}

\figcaption{BIMA Array response to source size for each comet. The upper panel is from our Comet LINEAR observations and the lower panel is from our Comet NEAT observations. The abscissa is cometocentric distance in arcseconds and the ordinate is the array recovery factor $r_i$. The region of the cometary coma that is dominated by collisions is labeled ``Collisional Zone''. The hatched area denotes the region where the transition between collisionally dominated and fluorescence dominated regions is. The region of the coma that is dominated by fluorescence is labeled ``Fluorescence Zone''. The solid line shows $r_i$ for each source size. The Array response to different source sizes shows that we are sampling the inner, collisionally dominated region of the coma.\label{fig:resp}}

\clearpage
\begin{figure}
\plotone{f1.eps}
\end{figure}
\clearpage
\begin{figure}
\epsscale{0.8}
\plotone{f2.eps}
\end{figure}

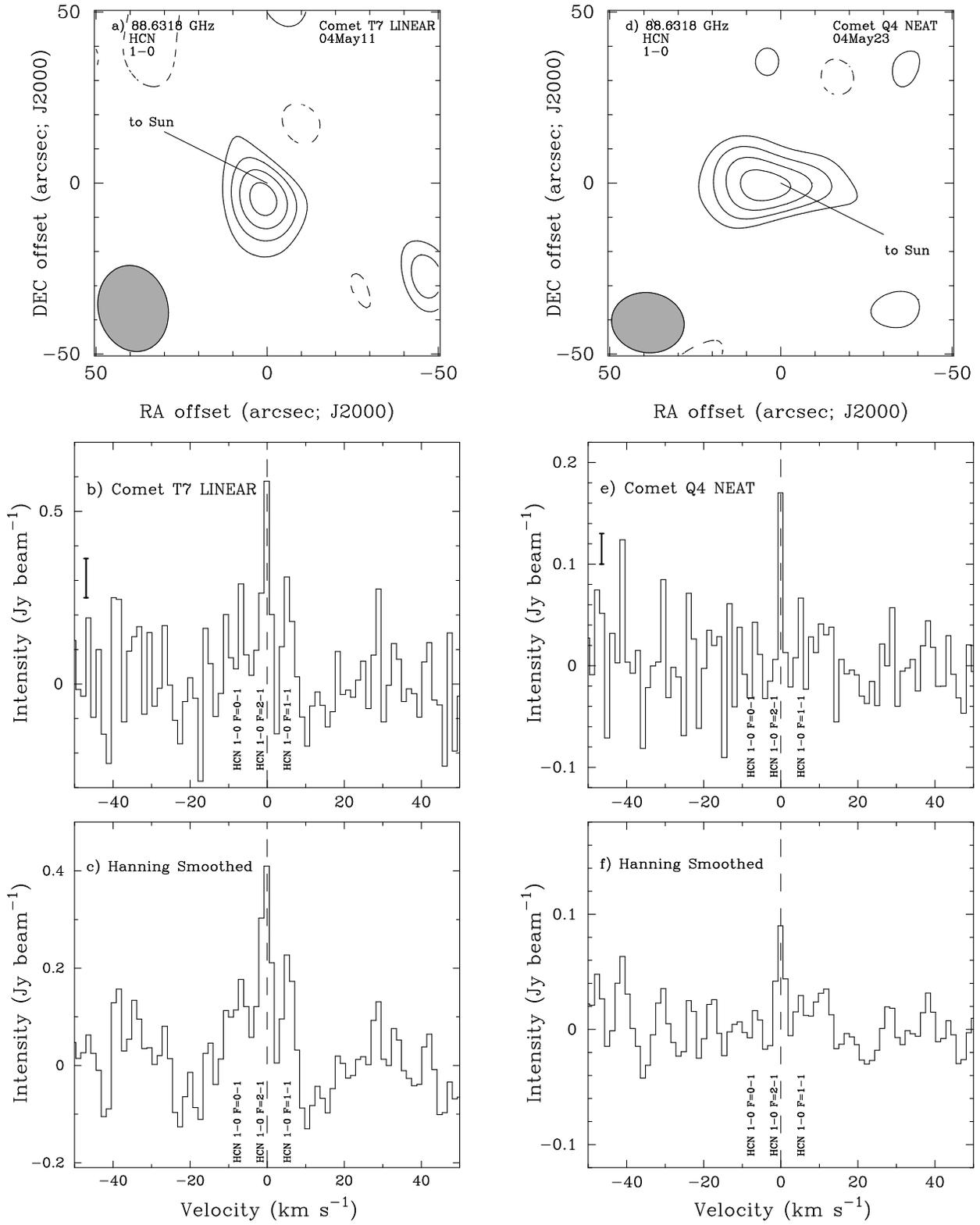
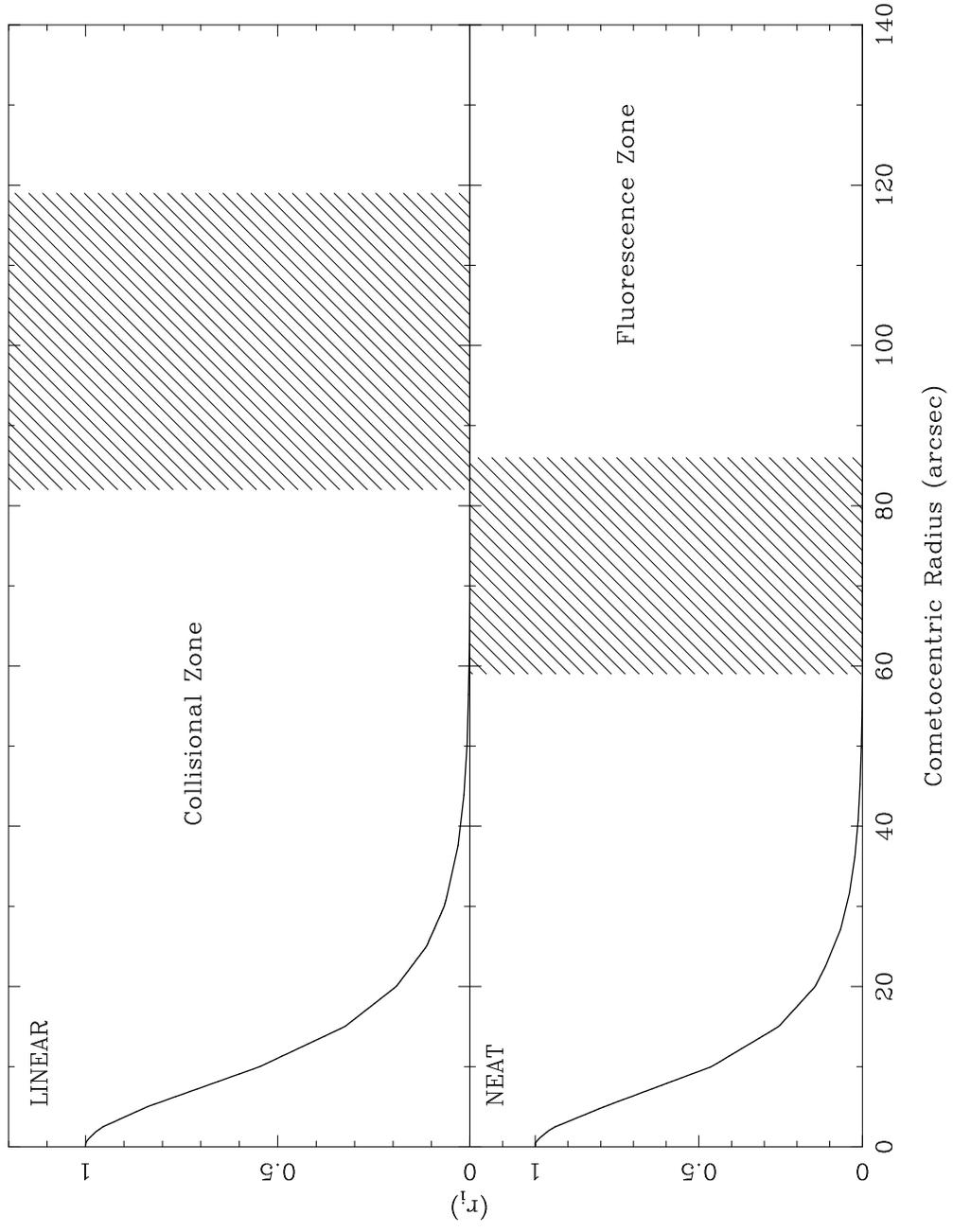
\end{document}